\documentclass[pra,twocolumn,superscriptaddress,
showpacs,preprintnumbers,amsmath,amssymb]{revtex4}

\usepackage{graphicx}
\usepackage{subfigure}
\usepackage{amsthm}
\usepackage{color}
\usepackage[all]{xy}
\usepackage{tikz}
\usepackage{dsfont}
\usetikzlibrary{positioning}
\usepackage{braket}

\newtheorem{Thm}{Theorem}

\newtheorem{Cor}[Thm]{Corollary}
\theoremstyle{definition}

\newtheorem{Rem}[Thm]{Remark}

\newcommand{\Tr}{\mathop{\mathrm{Tr}}\nolimits}

\begin{document}

\title{State Exchange with Quantum Side Information}

\author{Yonghae Lee}\email{yonghaelee@khu.ac.kr}
\affiliation{
Department of Mathematics and Research Institute for Basic Sciences,
Kyung Hee University, Seoul 02447, Korea}

\author{Ryuji Takagi}\email{rtakagi@mit.edu}
\affiliation{
Department of Physics and Center for Theoretical Physics, Massachusetts Institute of Technology,
Cambridge, Massachusetts 02139, USA}

\author{Hayata Yamasaki}\email{yamasaki@eve.phys.s.u-tokyo.ac.jp}
\affiliation{
Department of Physics, Graduate School of Science,
The University of Tokyo, 7--3--1 Hongo, Bunkyo-ku, Tokyo, Japan}

\author{Gerardo Adesso}\email{gerardo.adesso@nottingham.ac.uk}
\affiliation{
School of Mathematical Sciences
and Centre for the Mathematics and Theoretical Physics of Quantum Non-Equilibrium Systems,
University of Nottingham, University Park, Nottingham NG7 2RD, United Kingdom}

\author{Soojoon Lee}\email{level@khu.ac.kr}
\affiliation{
Department of Mathematics and Research Institute for Basic Sciences,
Kyung Hee University, Seoul 02447, Korea}
\affiliation{
School of Mathematical Sciences
and Centre for the Mathematics and Theoretical Physics of Quantum Non-Equilibrium Systems,
University of Nottingham, University Park, Nottingham NG7 2RD, United Kingdom}

\date{\today}

\begin{abstract}
We consider a quantum communication task between two users Alice and Bob,
in which Alice and Bob exchange their respective quantum information
by means of local operations and classical communication
assisted by shared entanglement.
Here,
we assume that Alice and Bob may have quantum side information,
not transferred,
and classical communication is free.
In this work,
we derive general upper and lower bounds for the least amount of entanglement
which is necessary to perfectly perform this task,
called the state exchange with quantum side information.
Moreover,
we show that the optimal entanglement cost can be negative
when Alice and Bob make use of their quantum side information.
We finally provide conditions on the initial state
for the state exchange with quantum side information
which give the exact optimal entanglement cost.
\end{abstract}

\maketitle

{\it Introduction.}---
In quantum information theory,
one of the most traditional research topics
has been source coding problems
of transmitting Alice's quantum information to Bob
under various situations, with paradigmatic examples including
Schumacher compression~\cite{S95} and quantum teleportation~\cite{BBCJPW93}.
A decade ago,
Oppenheim and Winter devised
a new type of a quantum communication task named
\emph{state exchange}~\cite{OW08} ---
in which Alice and Bob exchange their quantum information with each other
by means of local operations and classical communication (LOCC)
and shared entanglement ---
and they studied the least amount of entanglement consumed in the task
when free classical communication is allowed.

In the original state exchange task,
it is assumed that
both Alice and Bob do not have any quantum side information (QSI)
transferrable during the protocol.
On the other hand,
most quantum communication tasks,
including state merging~\cite{HOW05,HOW07} and state redistribution~\cite{DY08,YD09},
begin with the assumption that either
Alice or Bob has QSI.
For example,
in the  state merging task,
Bob can make use of his QSI for merging Alice's information to him,
and 
the minimum amount of entanglement needed for merging turns out to be exactly
given by the quantum conditional entropy~\cite{W13} conditioned on Bob's QSI.

\begin{figure}
\centering
\includegraphics[width=.9\linewidth,trim=0cm 0cm 0cm 0cm]{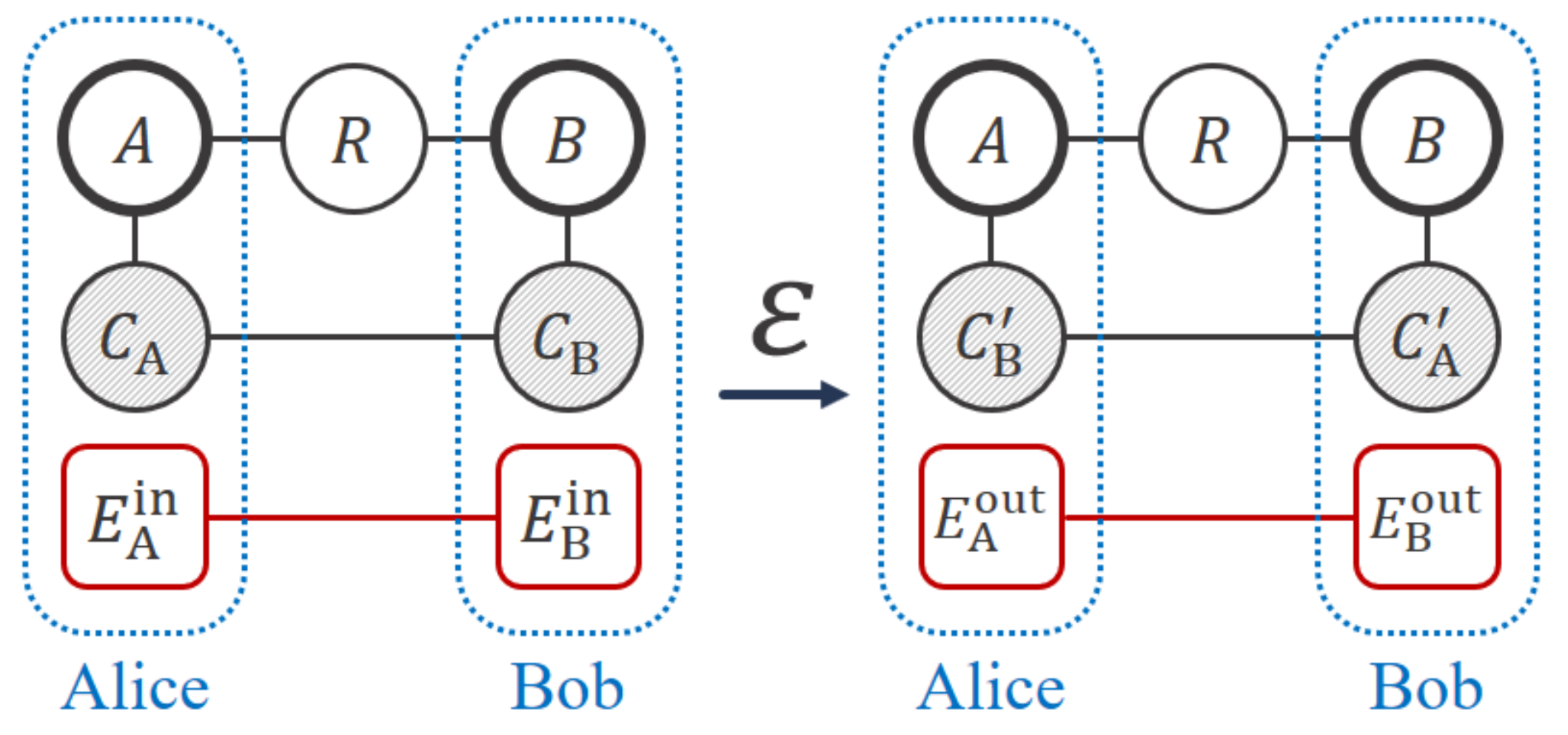}
\caption{
Illustration of state exchange protocol $\mathcal{E}$ with QSI. Starting from an initial  state $\ket{\psi}_{AC_{\mathrm{A}}BC_{\mathrm{B}}R}$ of Alice, Bob, and a referee ($R$),  Alice and Bob exchange their parts $C_{\mathrm{A}}$ and $C_{\mathrm{B}}$, exploiting their respective QSI $A$ and $B$. The ancillary systems $E_{\mathrm{A}}^{\mathrm{in}}$ and $E_{\mathrm{B}}^{\mathrm{in}}$ represent an initial entanglement consumed for the exchanging task,
while $E_{\mathrm{A}}^{\mathrm{out}}$ and $E_{\mathrm{B}}^{\mathrm{out}}$
indicate entanglement generated from the task.
}
\label{fig:SEwQSI}
\end{figure}

In this work
we generalize in the state exchange to
an exchanging task allowing Alice's and Bob's QSI,
which is called the \emph{state exchange with quantum side information}.
We consider three parties, Alice, Bob, and a referee ($R$),
sharing a pure initial state $\ket{\psi}\equiv\ket{\psi}_{AC_{\mathrm{A}}BC_{\mathrm{B}}R}$
as depicted in Fig.~\ref{fig:SEwQSI}.
The aim of Alice and Bob is
to exchange their quantum information $C_{\mathrm{A}}$ and $C_{\mathrm{B}}$,
while the referee does nothing.
To achieve their aim,
Alice and Bob make use of their QSI $A$ and $B$,
and they have additional systems
$E_{\mathrm{A}}^{\mathrm{in}}$, $E_{\mathrm{A}}^{\mathrm{out}}$
and $E_{\mathrm{B}}^{\mathrm{in}}$, $E_{\mathrm{B}}^{\mathrm{out}}$
for the use of entanglement resources.

Our main question can be formulated as follows:
{\em ``Does there exist a crucial difference in optimal strategies
between the tasks of state exchange with and without QSI?''}

To answer this question
we formally define the state exchange with QSI and its optimal entanglement cost
in the asymptotic scenario,
and then derive an upper bound for the optimal entanglement cost
by conceiving a two-step strategy based on the idea mentioned in Ref.~\cite{OW08}.
We show that in general
this strategy does not provide the optimal entanglement cost of the state exchange with QSI.
However
for a specific initial state of the state exchange with QSI,
the upper bound shows
that the optimal entanglement cost for the state exchange with QSI can be \emph{negative}, meaning that entanglement is in fact gained rather than consumed in the protocol.
This result is quite remarkable
since the optimal entanglement cost for the state exchange without QSI cannot be negative \cite{OW08}.
More importantly,
this implies
that the use of Alice's and Bob's QSI can significantly reduce
the optimal entanglement cost of the exchanging task.

We furthermore consider an idealized situation in which
the referee plays a more active role and can help Alice and Bob to exchange their information~\cite{OW08}.
By virtue of the referee's assistance,
it is possible for Alice and Bob to more efficiently perform the state exchange with QSI,
and this provides us with converse bounds on the optimal entanglement cost,
which are lower bounds for any achievable entanglement rate.
As an application of our bounds,
we present conditions on the initial state for the state exchange with QSI
such that the exact optimal entanglement cost can be obtained.

{\it State exchange with quantum side information.}---
In the task of state exchange $\mathcal{E}$ with QSI as described in Fig.~\ref{fig:SEwQSI},
the global initial state $\psi_i$ and the global final state $\psi_f$
are given by
\begin{equation}
\psi_i
=\psi
   \otimes{\Phi}_{E_{\mathrm{A}}^{\mathrm{in}}E_{\mathrm{B}}^{\mathrm{in}}}
\quad \mathrm{and} \quad
\psi_f
=\psi' 
   \otimes{\Phi}_{E_{\mathrm{A}}^{\mathrm{out}}E_{\mathrm{B}}^{\mathrm{out}}}, \nonumber
\end{equation}
where $\psi=\ket{\psi}\bra{\psi}$,
${\Phi}_{E_{\mathrm{A}}^{\mathrm{in}}E_{\mathrm{B}}^{\mathrm{in}}}$
and ${\Phi}_{E_{\mathrm{A}}^{\mathrm{out}}E_{\mathrm{B}}^{\mathrm{out}}}$
are pure maximally entangled states
with Schmidt rank $e^{\mathrm{in}}(\mathcal{E})$ and $e^{\mathrm{out}}(\mathcal{E})$,
respectively,
$
\psi'=\left(\mathds{1}_{ABR}
\otimes
\mathds{1}_{C_{\mathrm{A}}\to C_{\mathrm{A}}'}
\otimes
\mathds{1}_{C_{\mathrm{B}}\to C_{\mathrm{B}}'}
\right)(\psi)
$, and 
$C_{\mathrm{B}}'$ ($C_{\mathrm{A}}'$) is Alice's system (Bob's system)
with $\dim C_{\mathrm{B}}'=\dim C_{\mathrm{B}}$ ($\dim C_{\mathrm{A}}'=\dim C_{\mathrm{A}}$).
Then
a joint operation
\begin{equation}
\mathcal{E}:AC_{\mathrm{A}}E_{\mathrm{A}}^{\mathrm{in}}\otimes
BC_{\mathrm{B}}E_{\mathrm{B}}^{\mathrm{in}}
\longrightarrow AC_{\mathrm{B}}'E_{\mathrm{A}}^{\mathrm{out}}\otimes
BC_{\mathrm{A}}'E_{\mathrm{B}}^{\mathrm{out}} \nonumber
\end{equation}
is called \emph{state exchange with quantum side information of $\ket{\psi}$ with error $\varepsilon$},
if it consists of LOCC,
and satisfies
\begin{equation}
\left\|
\left(\mathcal{E}\otimes\mathds{1}_{R}\right)
\left(\psi_i\right)
-
\psi_f
\right\|_{1}
\le\varepsilon, \nonumber
\end{equation}
where
$\|\cdot\|_1$ is the trace norm~\cite{W13}.

Let us now consider $n$ independent and identically distributed copies
of $\ket{\psi}$,
say $\ket{\psi}^{\otimes n}$.
If $\mathcal{E}_n$ indicates a state exchange
with QSI of $\ket{\psi}^{\otimes n}$ with error $\varepsilon_n$,
then the resource rate
$\left(\log e^{\mathrm{in}}(\mathcal{E}_n)-\log e^{\mathrm{out}}(\mathcal{E}_n)\right)/n$
is called the \emph{entanglement rate} of the protocol.
If there is a sequence  $\{\mathcal{E}_n\}_{n\in\mathbb{N}}$ of state exchanges $\mathcal{E}_n$
with QSI of $\ket{\psi}^{\otimes n}$ with error $\varepsilon_n$
such that 
\begin{equation*}
\lim_{n\to\infty}
\frac{\log e^{\mathrm{in}}(\mathcal{E}_n)-\log e^{\mathrm{out}}(\mathcal{E}_n)}{n}
=e_\mathrm{r}, \nonumber \qquad %
\lim_{n\to\infty} \varepsilon_n=0, 
\end{equation*}
then the real number $e_\mathrm{r}$ is called
an \emph{achievable} entanglement rate for the state exchange with QSI
of $\ket{\psi}$. 
The smallest achievable entanglement rate defines 
the \emph{optimal entanglement cost} $e_{\mathrm{opt}}$ for the considered task.

{\it Merge-and-merge strategy.}---
We first present a \emph{merge-and-merge} strategy
which is motivated by the \emph{merge-and-send} protocol introduced in Ref.~\cite{OW08}.
The idea of this strategy is as follows.
Firstly,
Alice's part $C_{\mathrm{A}}$ is merged from Alice to Bob
by using $BC_{\mathrm{B}}$ as QSI.
After finishing merging $C_{\mathrm{A}}$,
Bob's part $C_{\mathrm{B}}$ is merged from Bob to Alice
by using Alice's QSI $A$
so that Alice's $C_{\mathrm{A}}$ and Bob's $C_{\mathrm{B}}$ are exchanged.
By using the exact formula of the entanglement cost for merging~\cite{O08,DY08,LL18},
we have that
the optimal entanglement costs of merging $C_{\mathrm{A}}$ and merging $C_{\mathrm{B}}$
are the quantum conditional entropies
$H(C_{\mathrm{A}}|BC_{\mathrm{B}})$ and $H(C_{\mathrm{B}}|A)$,
respectively,
so that the total entanglement cost is $H(C_{\mathrm{B}}|A)+H(C_{\mathrm{A}}|BC_{\mathrm{B}})$, where the quantum conditional entropy $H(X|Y)_{\rho}$ of a state $\rho_{XY}$
is defined by $H(XY)_{\rho}-H(Y)_{\rho}$, with $H(X)$ the von Neumann entropy~\cite{W13} of a state $\rho_X$.

From the merge-and-merge strategy,
we obtain the following upper bound
for the optimal entanglement cost of the state exchange with QSI.

\begin{Thm} \label{thm:ub}
The optimal entanglement cost $e_{\mathrm{opt}}$
for the state exchange with QSI of $\ket{\psi}$ 
is upper bounded by
\begin{equation}
e_{\mathrm{opt}} \le u(\psi)=\min\{ u_{1}(\psi), u_{2}(\psi) \}, \nonumber
\end{equation}
where $u_{1}(\psi)=H(C_{\mathrm{B}}|A)_{\psi}+H(C_{\mathrm{A}}|BC_{\mathrm{B}})_{\psi}$
and $u_{2}(\psi)=H(C_{\mathrm{A}}|B)_{\psi}+H(C_{\mathrm{B}}|AC_{\mathrm{A}})_{\psi}$.
\end{Thm}

Note that $u_2(\psi)$ in Theorem~\ref{thm:ub} can be obtained
by firstly merging Bob's part $C_{\mathrm{B}}$ to Alice.
We further refer the reader to Appendix~\ref{app:proof_ub}
for the rigorous proof of Theorem~\ref{thm:ub}
which fulfills the definition of achievability.

{\it Optimal strategy?}---
Since the merge-and-merge strategy is simple and intuitive,
one may guess that the strategy is optimal for any initial state of the exchanging task.
However,
the following example shows that
there can be a more effective strategy than the merge-and-merge one.
Let us consider a specific form of the initial state
\begin{equation}
\ket{\tilde{\psi}}_{AC_\mathrm{A}BC_\mathrm{B}R}
=\ket{\tilde{\phi}}_{AC_\mathrm{A}^1BC_\mathrm{B}^1R_1}\otimes\ket{\mathrm{GHZ}}_{{C_\mathrm{A}^2}{C_\mathrm{B}^2}R_2},
\label{eq:upper_gap_example}
\end{equation}
where systems $C_\mathrm{A}=C_\mathrm{A}^1C_\mathrm{A}^2$, $C_\mathrm{B}=C_\mathrm{B}^1C_\mathrm{B}^2$,
$R=R_1R_2$, 
$\ket{\tilde{\phi}}$ is an arbitrary state on the system $AC_\mathrm{A}^1BC_\mathrm{B}^1R_1$, and
\begin{equation}
\ket{\mathrm{GHZ}}_{{C_\mathrm{A}^2}{C_\mathrm{B}^2}R_2}=\frac{1}{\sqrt{d}}\sum_{k=0}^{d-1}\ket{kkk} \nonumber
\label{eq:GHZ}
\end{equation}
is the Greenberger-Horne-Zeilinger state~\cite{GHZ89} with $d\ge2$.

In order to exchange $C_\mathrm{A}$ and $C_\mathrm{B}$ in Eq.~(\ref{eq:upper_gap_example}),
it suffices for Alice and Bob to only consider the state exchange with QSI of $\ket{\tilde{\phi}}$,
since the state $\ket{\mathrm{GHZ}}$
on the parts $C_{\mathrm{A}}^2$ and $C_{\mathrm{B}}^2$ is symmetric.
Then by applying the merge-and-merge strategy on $\ket{\tilde{\phi}}$,
we obtain a tighter upper bound $\min\{ u_{1}(\tilde{\phi}), u_{2}(\tilde{\phi}) \}$
for the optimal entanglement cost for the state $\ket{\tilde{\psi}}$
in Eq.~(\ref{eq:upper_gap_example})
as follows:
\begin{equation}
\min\{ u_{1}(\tilde{\phi}), u_{2}(\tilde{\phi}) \}
=\min\{ u_{1}(\tilde{\psi}), u_{2}(\tilde{\psi}) \}- \log{d}.
\label{eq:3logd}
\end{equation}

From the relation between upper bounds in Eq.~(\ref{eq:3logd}),
we remark that
there can be an arbitrarily large gap between the optimal entanglement cost
and the upper bound in Theorem~\ref{thm:ub},
implying that the upper bound is not optimal in the general case.
This example also shows that
there exist tighter upper bounds for the optimal entanglement cost.
On this account, we argue that the optimal strategy for state exchange with QSI is generally nontrivial.

{\it Converse bounds.}---
As in the state exchange \emph{without} QSI~\cite{OW08},
we can imagine that the referee holds the reference $R$,
and is ideally allowed to assist Alice and Bob in the following way,
which is here called the \emph{$R$-assisted state exchange with QSI}.
The referee first divides their part $R$ into two parts $E$ and $V$
by using a quantum channel $\mathcal{N}$ from $R$ to $V$
whose complementary channel $\mathcal{N}^c$ is from $R$ to $E$~\cite{W13}.
Next, the referee sends the states $\rho_V=\mathcal{N}(\rho_R)$ and $\rho_E=\mathcal{N}^c(\rho_R)$
to Alice and Bob, respectively.
Then the initial state $\ket{\psi}$ becomes $\ket{\psi}_{AC_\mathrm{A}VBC_\mathrm{B}E}$,
where Alice and Bob hold $AC_\mathrm{A}V$ and $BC_\mathrm{B}E$, respectively.
Alice and Bob now perform the state exchange with QSI of the state $\ket{\psi}_{AC_\mathrm{A}VBC_\mathrm{B}E}$.

For each $n$, let $\mathcal{E}^R_n$ be a state exchange with QSI
of $\ket{\psi}^{\otimes n}$ with error $\varepsilon_n$,
and $E^{\mathrm{bef}}_n$ and $E^{\mathrm{aft}}_n$ be
total amounts of entanglement between Alice and Bob
before and after the state exchange with QSI, respectively.
Then they can be expressed as
$E^{\mathrm{bef}}_n=n H(AC_{\mathrm{A}}V)+\log e^{\mathrm{in}}(\mathcal{E}^R_n)$
and
$E^{\mathrm{aft}}_n=n H(AC_{\mathrm{B}}V)+\log e^{\mathrm{out}}(\mathcal{E}^R_n)$.
Since the total entanglement between Alice and Bob cannot increase
under LOCC~\cite{BDSW96},
we have $E^{\mathrm{bef}}_n \ge E^{\mathrm{aft}}_n$,
that is,
\begin{equation}
\log e^{\mathrm{in}}(\mathcal{E}^R_n)-\log e^{\mathrm{out}}(\mathcal{E}^R_n)
\ge nH(AC_{\mathrm{B}}V)-nH(AC_{\mathrm{A}}V). \nonumber
\end{equation}
Let $e_{\mathrm{opt}}^{R}$ be the optimal entanglement cost
for the $R$-assisted state exchange with QSI,
then 
\begin{equation}
\max_{\mathcal{N}}[H(AC_{\mathrm{B}}V)-H(AC_{\mathrm{A}}V)]\le e_{\mathrm{opt}}^{R}. \nonumber
\end{equation}
Since any state exchange with QSI can be considered as an $R$-assisted state exchange with QSI (in which the referee trivially does nothing), it holds that 
$e_{\mathrm{opt}}^R\le e_{\mathrm{opt}}$.
This leads us to the following theorem.

\begin{Thm} \label{thm:lb}
The optimal entanglement cost $e_{\mathrm{opt}}$
for the state exchange with QSI of $\ket{\psi}$
is lower bounded by
\begin{equation} \label{eq:max_lb}
l(\psi)
=\max_{\mathcal{N}}[H(AC_{\mathrm{B}}V)_{\mathcal{N}(\psi)}
-H(AC_{\mathrm{A}}V)_{\mathcal{N}(\psi)}]
\le e_{\mathrm{opt}}, \nonumber
\end{equation}
where the maximum is taken over all quantum channels $\mathcal{N}:R\longrightarrow V$.
\end{Thm}

In general,
it is not easy to calculate the converse bound in Theorem~\ref{thm:lb},
since it involves an optimization over all quantum channels.
However,
if the referee sends the whole part $R$ to either Alice or Bob
without dividing $R$ in Theorem~\ref{thm:lb},
then we obtain the following \emph{computable} converse bound:

\begin{Cor}\label{cor:lb}
For the state exchange with QSI of $\ket{\psi}$,
the optimal entanglement cost $e_{\mathrm{opt}}$
satisfies
\begin{equation}
\max\{ l_1(\psi), l_2(\psi) \}\le e_{\mathrm{opt}}, \nonumber
\end{equation}
where $l_1(\psi)=H(AC_\mathrm{B})_{\psi}-H(AC_\mathrm{A})_{\psi}$
and $l_2(\psi)=H(BC_\mathrm{A})_{\psi}-H(BC_\mathrm{B})_{\psi}$.
\end{Cor}

By using the continuity of the von Neumann entropy~\cite{F73,A07},
we can directly show that $l_1(\psi)$ and $l_2(\psi)$ in Corollary~\ref{cor:lb} are lower bounds
to the optimal entanglement cost for the state exchange with QSI of $\ket{\psi}$.
The proof of Corollary~\ref{cor:lb}
can be found in Appendix~\ref{app:proof_lb}.

{\it Large gap between converse bounds.}---
It is obvious that
the lower bound 
presented in Corollary~\ref{cor:lb}
is less tight than the one in Theorem~\ref{thm:lb}.
Interestingly,
the gap between these two converse bounds can be arbitrarily large.
To this end,
let us consider the initial state
\begin{equation}
\ket{\bar{\psi}}_{AC_\mathrm{A}BC_\mathrm{B}R}
=\ket{\Phi}_{AR_{A}}
\otimes\ket{\Phi}_{C_\mathrm{A}R_{C_\mathrm{A}}}\otimes\ket{\Phi}_{BR_{B}}
\otimes\ket{\Phi}_{C_\mathrm{B}R_{C_\mathrm{B}}}, 
\label{eq:gap_example}
\end{equation}
where the reference system $R$ consists of the four subsystems $R_{A}$, $R_{C_\mathrm{A}}$, $R_{B}$ and $R_{C_\mathrm{B}}$,
and $\ket{\Phi}$ is a maximally entangled state on the corresponding bipartite system $SR_S$
with $\dim S = \dim R_S$ for $S=A$, $B$, $C_\mathrm{A}$ and $C_\mathrm{B}$.
Then we can readily see that
\begin{equation}
l_1(\bar{\psi})
=H(C_\mathrm{B})_{\bar{\psi}}-H(C_\mathrm{A})_{\bar{\psi}}
=-l_2(\bar{\psi}). \nonumber
\label{eq:example_lbs}
\end{equation}
On the other hand,
if a channel $\bar{\mathcal{N}}$ is given by $\rho_R\mapsto \rho_{R_{A}R_{C_\mathrm{A}}}$,
that is, $V=R_\mathrm{A}R_{C_\mathrm{A}}$,
then we obtain
\begin{eqnarray}
l(\bar{\psi})
&\ge&H(AC_{\mathrm{B}}V)_{\bar{\mathcal{N}}(\bar{\psi})}
    -H(AC_{\mathrm{A}}V)_{\bar{\mathcal{N}}(\bar{\psi})} \nonumber \\
&=&H(AC_{\mathrm{B}}R_{A}R_{C_\mathrm{A}})_{\bar{\psi}}
  -H(AC_{\mathrm{A}}R_{A}R_{C_\mathrm{A}})_{\bar{\psi}} \nonumber \\
&=&H(C_\mathrm{A})_{\bar{\psi}}+H(C_\mathrm{B})_{\bar{\psi}}, \nonumber
\label{eq:gap_example_ec}
\end{eqnarray}
which means that
the converse bound $l(\psi)$ in Theorem~\ref{thm:lb} can be arbitrarily larger
than $\max\{l_1(\psi),l_2(\psi)\}$ in Corollary~\ref{cor:lb}
for the class of initial states in Eq.~(\ref{eq:gap_example}).

{\it Optimal entanglement cost can be negative.}---
We finally address the crucial question: Can the optimal entanglement cost for state exchange with QSI be {\em negative}?
First of all, let us remark that 
the optimal entanglement cost for state exchange {\em without}
QSI of $\ket{\psi}_{C_{\mathrm{A}}C_{\mathrm{B}}R}$ {\em cannot} be negative~\cite{OW08}.
If the optimal cost was negative,
then Alice and Bob could generate as much entanglement as they need
by repeatedly exchanging their state.
This contradicts the basic requirement that the amount of entanglement cannot increase by LOCC~\cite{VPRK97}.

\begin{figure}
\centering
\includegraphics[width=.95\linewidth,trim=0cm 0cm 0cm 0cm]{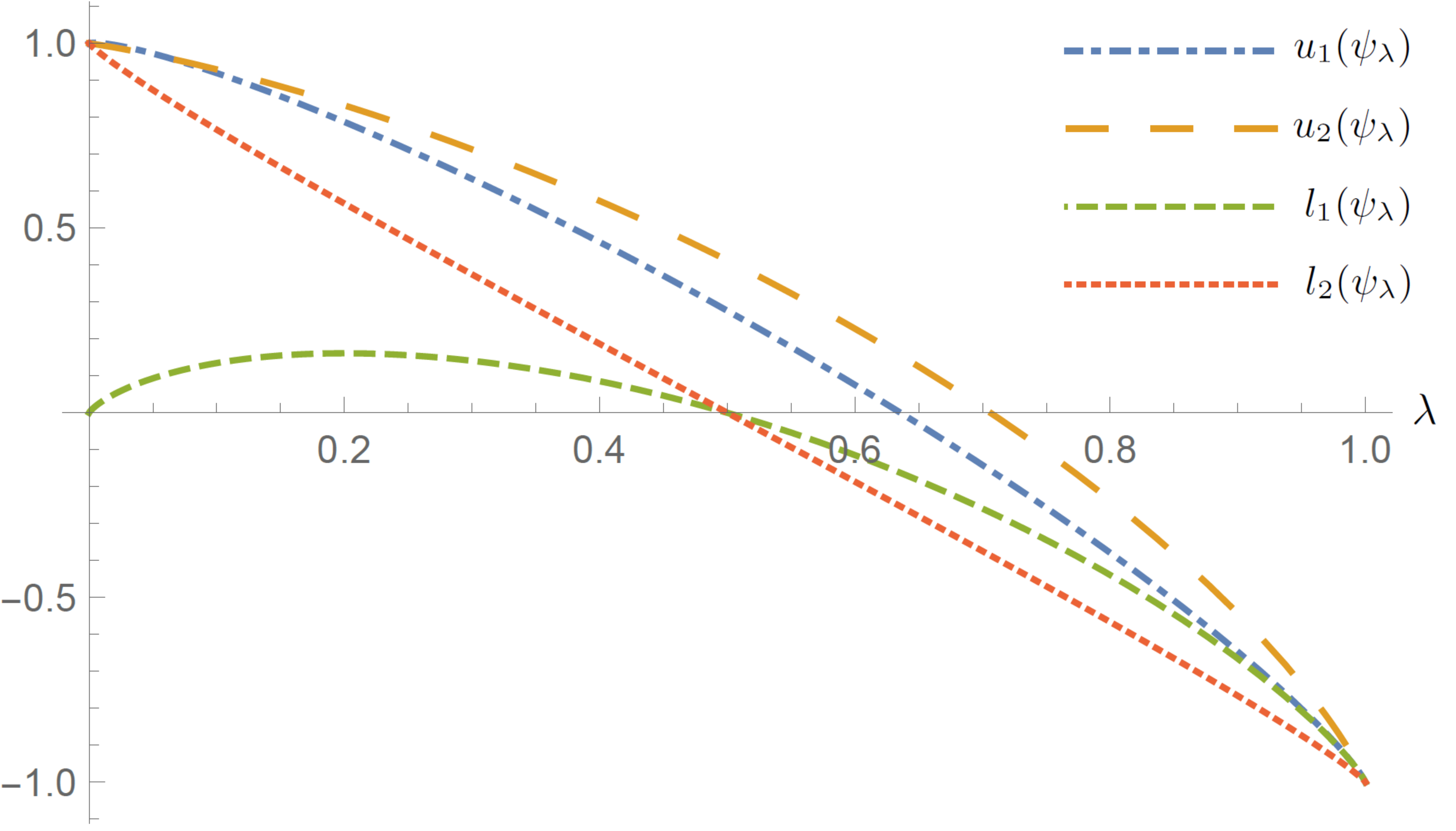}
\caption{
Upper bounds $u_1(\psi_{\lambda})$, $u_2(\psi_{\lambda})$
and lower bounds $l_1(\psi_{\lambda})$, $l_2(\psi_{\lambda})$
to the optimal entanglement cost $e_{\mathrm{opt}}$
for the specific initial state $\ket{\psi_{\lambda}}$ of Eq.~(\ref{eq:canbenegative}) with $0\le\lambda\le1$.
}
\label{fig:graph}
\end{figure}

However, quite
remarkably,
the optimal entanglement cost $e_{\mathrm{opt}}$
for the state exchange {\em with} QSI of $\ket{\psi}$ 
{\em can} be negative. This is readily seen
since the upper bounds $u_1$ or $u_2$ in Theorem~\ref{thm:ub} can be negative.
For example,
$e_{\mathrm{opt}}$ is negative
for the initial state
\begin{eqnarray}
\ket{\psi_{\lambda}}_{AC_{\mathrm{A}}BC_{\mathrm{B}}R}
&=&\sqrt{\frac{\lambda}{2}}\ket{00000}
+\sqrt{\frac{1-\lambda}{2}}\ket{10011} \nonumber \\
&+&\sqrt{\frac{\lambda}{2}}\ket{01100}
+\sqrt{\frac{1-\lambda}{2}}\ket{01010} 
\label{eq:canbenegative}
\end{eqnarray}
with $\lambda\ge0.65$, as seen in Fig.~\ref{fig:graph}.
Furthermore,
this example shows that,
in the state exchange with QSI, 
the optimal entanglement cost can be generally reduced
by exploiting the QSI $AB$ for the exchanging task. This reveals the prominent role of the QSI for such a quantum communication primitive.

At this point
we remark that the negativity of the optimal entanglement cost
for the state exchange with QSI does not lead to a contradiction as follows.
Let $e_{\mathrm{opt}}^{\mathrm{1st}}$ be the optimal entanglement cost
for a state exchange with QSI of the initial state $\ket{\psi}$,
and let $e_{\mathrm{opt}}^{\mathrm{2nd}}$ be the optimal entanglement cost
for a state exchange with QSI of the exchanged state $\ket{\psi'}$.
Then from Corollary~\ref{cor:lb},
\begin{equation}
e_{\mathrm{opt}}^{\mathrm{1st}}\ge l_1(\psi)
\quad \mathrm{and} \quad
e_{\mathrm{opt}}^{\mathrm{2nd}}\ge l_1(\psi')=-l_1(\psi). \nonumber
\end{equation}
So in this case we have
the inequality $e_{\mathrm{opt}}^{\mathrm{1st}}+e_{\mathrm{opt}}^{\mathrm{2nd}}\ge0$.
This shows that
the total amount of entanglement generated from repeated state exchange protocols with QSI
does not repeatedly increase
although the entanglement cost can be negative
in an individual instance of the protocol.

{\it Optimal entanglement costs for some special cases.}---
We now provide several conditions
which allow us to compute the exact optimal entanglement cost $e_{\mathrm{opt}}$
for the state exchange with QSI of $\ket{\psi}$.
In fact,
the merge-and-merge strategy is optimal under these conditions.

\begin{Cor} \label{cor:optimal_condition}
Let $e_{\mathrm{opt}}$ be the optimal entanglement cost
of the state exchange with QSI of $\ket{\psi}\equiv\ket{\psi}_{AC_{\mathrm{A}}BC_{\mathrm{B}}R}$.

(i) 
The following conditions on $\ket{\psi}$ give the exact optimal entanglement costs:
\begin{eqnarray} \label{eq:CMIconditions}
I(R;C_\mathrm{A}|A)_{\psi}=0 &\Longleftrightarrow& e_{\mathrm{opt}}=u_1(\psi)=l_1(\psi), \nonumber \\
I(R;C_\mathrm{A}|B)_{\psi}=0 &\Longleftrightarrow& e_{\mathrm{opt}}=u_2(\psi)=l_1(\psi), \nonumber \\
I(R;C_\mathrm{B}|A)_{\psi}=0 &\Longleftrightarrow& e_{\mathrm{opt}}=u_1(\psi)=l_2(\psi), \nonumber \\
I(R;C_\mathrm{B}|B)_{\psi}=0 &\Longleftrightarrow& e_{\mathrm{opt}}=u_2(\psi)=l_2(\psi), \nonumber
\end{eqnarray}
where $I(X;Y|Z)_{\rho}$ indicates the quantum conditional mutual information (QCMI)
of a quantum state $\rho_{XYZ}$,
and $u_{1}(\psi)$, $u_{2}(\psi)$, $l_1(\psi)$, and $l_2(\psi)$
are given in Theorem~\ref{thm:ub} and Corollary~\ref{cor:lb}.

(ii) 
There exists a quantum channel $\mathcal{N}:R\longrightarrow V$
such that
$ 
I(C_{\mathrm{B}}:V|A)_{\mathcal{N}(\psi)}
=I(C_{\mathrm{A}}:E|AV)_{\mathcal{N}(\psi)}=0
$ 
if and only if $e_{\mathrm{opt}}=u_1(\psi)=l(\psi)$,
where $l(\psi)$ is in Theorem~\ref{thm:lb}.
Similarly,
there exists 
$\mathcal{N}:R\longrightarrow V$
such that
$ 
I(C_{\mathrm{A}}:E|B)_{\mathcal{N}(\psi)}
=I(C_{\mathrm{B}}:V|BE)_{\mathcal{N}(\psi)}=0
$ 
if and only if $e_{\mathrm{opt}}=u_2(\psi)=l(\psi)$.

(iii) 
Let $\ket{\hat{\psi}}_{AC_{\mathrm{A}}BC_{\mathrm{B}}}$ be
a pure initial state shared by Alice and Bob (with no referee), then for the state exchange with QSI of $\ket{\hat{\psi}}_{AC_{\mathrm{A}}BC_{\mathrm{B}}}$ one has 
$ 
e_{\mathrm{opt}}=H(AC_{\mathrm{B}})_{\hat{\psi}}-H(AC_{\mathrm{A}})_{\hat{\psi}}
$. 
\end{Cor}

By combining the aforementioned upper and lower bounds,
the conditions for the exact optimal cost in Corollary~\ref{cor:optimal_condition}
are directly obtained. 
We remark that
there are no general implications among the four QCMI conditions
in Corollary~\ref{cor:optimal_condition} (i),
that is,
there exists an initial state which only satisfies one of these QCMI conditions.
We presents related examples in Appendix~\ref{app:example}.

{\it Conclusion.}---
In this work,
we have considered the state exchange with QSI as a fundamental quantum communication task,
and have provided the formal descriptions for the protocol and its optimal entanglement cost.
We have derived upper and lower bounds to the optimal entanglement cost.
From these bounds,
we have exactly evaluated the optimal entanglement cost for several special classes of states, including all pure bipartite states.
Furthermore,
we have shown that the optimal entanglement cost
for the state exchange with QSI can be negative.
This is at striking variance with the state exchange without QSI,
whose entanglement cost is always nonnegative.

By replacing classical communication with quantum communication,
we can consider a fully quantum version of the state exchange with QSI
of $\ket{\psi}_{AC_{\mathrm{A}}BC_{\mathrm{B}}R}$.
Similar to the idea of Theorem~\ref{thm:ub},
this task can be performed
by applying the state redistribution protocol~\cite{DY08,YD09} twice.
For example,
if the part $C_{\mathrm{A}}$ is firstly redistributed from Alice to Bob in this strategy,
then its achievable rates $E_\mathrm{r}$ and $Q_\mathrm{r}$ for ebits and qubit channels
are given by
\begin{eqnarray}
E_\mathrm{r}&=&\frac{1}{2}[l_1(\psi)+l_2(\psi)], \nonumber \\
Q_\mathrm{r}
&=&\frac{1}{2}u_1(\psi)
+\frac{1}{2}[H(C_{\mathrm{A}}|A)_{\psi}+H(C_{\mathrm{B}}|BC_{\mathrm{A}})_{\psi}], \nonumber
\end{eqnarray}
where $u_{1}(\psi)$, $l_1(\psi)$, and $l_2(\psi)$
are in Theorem~\ref{thm:ub} and Corollary~\ref{cor:lb}.
However,
in this case
the achievable region of a resource pair $(E_\mathrm{r},Q_\mathrm{r})$
is completely unknown.

To the best of our knowledge,
a protocol exchanging Alice's and Bob's information in a single step
has not been known,
and so in this work we have considered the merge-and-merge strategy,
in order to obtain achievable entanglement rates.
Hence it would be very meaningful to devise
one such a direct exchanging protocol.
Moreover,
recent results for one-shot quantum state merging~\cite{YM18}
and implementing bipartite unitaries~\cite{WSM17} may be useful
to figure out novel strategies
which can provide tighter achievable bounds than those in Theorem~\ref{thm:ub}.

Finally,
we expect
that studying variations on the state exchange with QSI
makes quantum information theory richer.
For example,
one can assume that
Alice and Bob can consume noisy resources~\cite{DHW04,DHW08} instead of noiseless resources,
or that Alice or Bob is additionally allowed to make use of
a local resource,
such as maximally coherent states~\cite{BCP14,SSDBA15,SCRBWL16},
as in the incoherent state merging~\cite{SCRBWL16}
and the incoherent state redistribution~\cite{AJS18}.
Exploring these avenues deserves further investigation.

\begin{acknowledgments}
We would like to thank Ludovico Lami, Bartosz Regula,
and Mario Berta for fruitful discussion.
This research was supported by the Basic Science Research Program
through the National Research Foundation of Korea (NRF)
funded by the Ministry of Science and ICT (NRF-2016R1A2B4014928).
R.T. acknowledges support from the Takenaka scholarship foundation.
G.A. acknowledges support from the ERC Starting Grant GQCOP (Grant Agreement No.~637352).
\end{acknowledgments}

\bibliography{SEwQSI}

\clearpage
\begin{widetext}
\newpage
\appendix
\setcounter{equation}{0}
\setcounter{page}{1}


\section{Proof of Theorem~\ref{thm:ub}} \label{app:proof_ub}

We first show that $u_{1}$ is achievable.
From the definition of the optimal costs for the state merging with QSI~\cite{LL18},
there are two sequences $\{\mathcal{F}^{\rightarrow}_{n}\}_{n\in\mathbb{N}}$
and $\{\mathcal{S}^{\leftarrow}_{n}\}_{n\in\mathbb{N}}$.
To be specific,
an element $\mathcal{F}^{\rightarrow}_{n}$
of the first sequence $\{\mathcal{F}^{\rightarrow}_{n}\}_{n\in\mathbb{N}}$,
\begin{equation*}
\mathcal{F}^{\rightarrow}_{n}
:
\left(
A^{\otimes n}C_{\mathrm{A}}^{\otimes n}F_{\mathrm{A}_n}^{\mathrm{in}}
\right)
\otimes
\left(
B^{\otimes n}C_{\mathrm{B}}^{\otimes n}F_{\mathrm{B}_n}^{\mathrm{in}}
\right)
\longrightarrow
\left(
A^{\otimes n}F_{\mathrm{A}_n}^{\mathrm{out}}
\right)
\otimes
\left(
C_{\mathrm{A}}'^{\otimes n}B^{\otimes n}C_{\mathrm{B}}^{\otimes n}F_{\mathrm{B}_n}^{\mathrm{out}}
\right),
\end{equation*}
is the state merging with QSI of $\ket{\psi}^{\otimes n}$
with error $\varepsilon_n^{\rightarrow}$
which is a LOCC operation satisfying
\begin{equation*}
\left\|
(\mathcal{F}^{\rightarrow}_{n}\otimes\mathds{1}_{R}^{\otimes n})
(\ket{\psi}^{\otimes n}
\otimes\ket{\Phi}_{F_{\mathrm{A}_n}^{\mathrm{in}}F_{\mathrm{B}_n}^{\mathrm{in}}})
-
(
\ket{\psi_\rightarrow}^{\otimes n}
\otimes\ket{\Phi}_{F_{\mathrm{A}_n}^{\mathrm{out}}F_{\mathrm{B}_n}^{\mathrm{out}}}
)
\right\|_{1}
\le\varepsilon_n^{\rightarrow}
\end{equation*}
where
$C_{\mathrm{A}}'$ is Bob's system with $\dim C_{\mathrm{A}}'=\dim C_{\mathrm{A}}$,
$\ket{\psi_\rightarrow}$ is a target state defined as
$\left(\mathds{1}_{ABC_{\mathrm{B}}R}
\otimes\mathds{1}_{C_{\mathrm{A}}\to C_{\mathrm{A}}'}\right)\ket{\psi}$,
and $\ket{\Phi}_{F_{\mathrm{A}_n}^{\mathrm{in}}F_{\mathrm{B}_n}^{\mathrm{in}}}$
and $\ket{\Phi}_{F_{\mathrm{A}_n}^{\mathrm{out}}F_{\mathrm{B}_n}^{\mathrm{out}}}$
are maximally entangled states
with Schmidt rank $e^{\mathrm{in}}(\mathcal{F}^{\rightarrow}_{n})$
and $e^{\mathrm{out}}(\mathcal{F}^{\rightarrow}_{n})$,
respectively.
An element $\mathcal{S}^{\leftarrow}_{n}$
of the second sequence $\{\mathcal{S}^{\leftarrow}_{n}\}_{n\in\mathbb{N}}$,
\begin{equation*}
\mathcal{S}^{\leftarrow}_{n}:
\left(
A^{\otimes n}S_{\mathrm{A}_n}^{\mathrm{in}}
\right)
\otimes
\left(
C_{\mathrm{A}}'^{\otimes n}B^{\otimes n}C_{\mathrm{B}}^{\otimes n}S_{\mathrm{B}_n}^{\mathrm{in}}
\right)
\longrightarrow
\left(
A^{\otimes n}C_{\mathrm{B}}'^{\otimes n}S_{\mathrm{A}_n}^{\mathrm{out}}
\right)
\otimes
\left(
C_{\mathrm{A}}'^{\otimes n}B^{\otimes n}S_{\mathrm{B}_n}^{\mathrm{out}}
\right),
\end{equation*}
is the state merging with QSI of $\ket{\psi_\rightarrow}^{\otimes n}$
with error $\varepsilon_n^{\leftarrow}$
which satisfies
\begin{equation*}
\left\|
(\mathcal{S}^{\leftarrow}_{n}\otimes\mathds{1}_{R}^{\otimes n})
(\ket{\psi_\rightarrow}^{\otimes n}
\otimes\ket{\Phi}_{S_{\mathrm{A}_n}^{\mathrm{in}}S_{\mathrm{B}_n}^{\mathrm{in}}})
-
(
\ket{\psi_\rightleftarrows}^{\otimes n}
\otimes\ket{\Phi}_{S_{\mathrm{A}_n}^{\mathrm{out}}S_{\mathrm{B}_n}^{\mathrm{out}}}
)
\right\|_{1}
\le\varepsilon_n^{\leftarrow}
\end{equation*}
where
$C_{\mathrm{B}}'$ is Alice's system with $\dim C_{\mathrm{B}}'=\dim C_{\mathrm{B}}$,
$\ket{\psi_\rightleftarrows}$ is a target state defined as
$\left(\mathds{1}_{AC_{\mathrm{A}}'BR}
\otimes\mathds{1}_{C_{\mathrm{B}}\to C_{\mathrm{B}}'}\right)\ket{\psi_\rightarrow}$,
and $\ket{\Phi}_{S_{\mathrm{A}_n}^{\mathrm{in}}S_{\mathrm{B}_n}^{\mathrm{in}}}$
and $\ket{\Phi}_{S_{\mathrm{A}_n}^{\mathrm{out}}S_{\mathrm{B}_n}^{\mathrm{out}}}$
are maximally entangled states
with Schmidt rank $e^{\mathrm{in}}(\mathcal{S}^{\leftarrow}_{n})$
and $e^{\mathrm{out}}(\mathcal{S}^{\leftarrow}_{n})$,
respectively.
The two sequences also satisfy
\begin{eqnarray*}
\lim_{n\to\infty}
\frac{\log e^{\mathrm{in}}(\mathcal{F}^{\rightarrow}_{n})
-\log e^{\mathrm{out}}(\mathcal{F}^{\rightarrow}_{n})}{n}
&=&H(C_{\mathrm{A}}|BC_{\mathrm{B}}), \\
\lim_{n\to\infty}
\frac{\log e^{\mathrm{in}}(\mathcal{S}^{\leftarrow}_{n})
-\log e^{\mathrm{out}}(\mathcal{S}^{\leftarrow}_{n})}{n}
&=&H(C_{\mathrm{B}}|A), \\
\lim_{n\to\infty} \varepsilon_n^{\rightarrow}=\lim_{n\to\infty} \varepsilon_n^{\leftarrow}&=&0.
\end{eqnarray*}

Let us now consider a sequence $\{\mathcal{E}^{\rightleftarrows}_n\}_{n\in\mathbb{N}}$ defined as
\begin{equation*}
\mathcal{E}^{\rightleftarrows}_n=
\begin{cases}
\hat{\mathcal{S}}^{\leftarrow}_{n}\circ\mathcal{F}^{\rightarrow}_{n}
&\text{if $e^{\mathrm{out}}(\mathcal{F}^{\rightarrow}_{n})
          \ge e^{\mathrm{in}}(\mathcal{S}^{\leftarrow}_{n})$} \\
\mathcal{S}^{\leftarrow}_{n}\circ\tilde{\mathcal{F}}^{\rightarrow}_{n}
&\text{otherwise},
\end{cases}
\end{equation*}
where
$\hat{\mathcal{S}}^{\leftarrow}_{n}
=\mathcal{S}^{\leftarrow}_{n}
\otimes
\mathds{1}_{\hat{E}_{\mathrm{A}_n}\hat{E}_{\mathrm{B}_n}}$
and
$\tilde{\mathcal{F}}^{\rightarrow}_{n}
=\mathcal{F}^{\rightarrow}_{n}
\otimes
\mathds{1}_{\tilde{E}_{\mathrm{A}_n}\tilde{E}_{\mathrm{B}_n}}$
with
$\dim \hat{E}_{\mathrm{A}_n}
=\dim \hat{E}_{\mathrm{B}_n}
=e^{\mathrm{out}}(\mathcal{F}^{\rightarrow}_{n})/e^{\mathrm{in}}(\mathcal{S}^{\leftarrow}_{n})$
and
$\dim \tilde{E}_{\mathrm{A}_n}
=\dim \tilde{E}_{\mathrm{B}_n}
=e^{\mathrm{in}}(\mathcal{S}^{\leftarrow}_{n})/e^{\mathrm{out}}(\mathcal{F}^{\rightarrow}_{n})$.
If
$e^{\mathrm{out}}(\mathcal{F}^{\rightarrow}_{n})\ge e^{\mathrm{in}}(\mathcal{S}^{\leftarrow}_{n})$
then
\begin{eqnarray*}
& &\left\|
(\mathcal{E}^{\rightleftarrows}_n\otimes\mathds{1}_{R}^{\otimes n})
(\ket{\psi}^{\otimes n}
\otimes\ket{\Phi}_{F_{\mathrm{A}_n}^{\mathrm{in}}F_{\mathrm{B}_n}^{\mathrm{in}}})
-
(
\ket{\psi_\rightleftarrows}^{\otimes n}
\otimes\ket{\Phi}_{S_{\mathrm{A}_n}^{\mathrm{out}}S_{\mathrm{B}_n}^{\mathrm{out}}}
\otimes\ket{\Phi}_{\hat{E}_{\mathrm{A}_n}\hat{E}_{\mathrm{B}_n}}
)
\right\|_{1}  \\
&&\le
\left\|
((\hat{\mathcal{S}}^{\leftarrow}_{n}\circ\mathcal{F}^{\rightarrow}_{n})
\otimes\mathds{1}_{R}^{\otimes n})
(\ket{\psi}^{\otimes n}
\otimes\ket{\Phi}_{F_{\mathrm{A}_n}^{\mathrm{in}}F_{\mathrm{B}_n}^{\mathrm{in}}})
-
(\hat{\mathcal{S}}^{\leftarrow}_{n}\otimes\mathds{1}_{R}^{\otimes n})
(\ket{\psi_\rightarrow}^{\otimes n}
\otimes\ket{\Phi}_{F_{\mathrm{A}_n}^{\mathrm{out}}F_{\mathrm{B}_n}^{\mathrm{out}}})
\right\|_{1}  \\
& &~+\left\|
(\hat{\mathcal{S}}^{\leftarrow}_{n}\otimes\mathds{1}_{R}^{\otimes n})
(\ket{\psi_\rightarrow}^{\otimes n}
\otimes\ket{\Phi}_{F_{\mathrm{A}_n}^{\mathrm{out}}F_{\mathrm{B}_n}^{\mathrm{out}}})
-
(
\ket{\psi_\rightleftarrows}^{\otimes n}
\otimes\ket{\Phi}_{S_{\mathrm{A}_n}^{\mathrm{out}}S_{\mathrm{B}_n}^{\mathrm{out}}}
\otimes\ket{\Phi}_{\hat{E}_{\mathrm{A}_n}\hat{E}_{\mathrm{B}_n}}
)
\right\|_{1}  \\
&&\le
\left\|
(\mathcal{F}^{\rightarrow}_{n}\otimes\mathds{1}_{R}^{\otimes n})
(\ket{\psi}^{\otimes n}
\otimes\ket{\Phi}_{F_{\mathrm{A}_n}^{\mathrm{in}}F_{\mathrm{B}_n}^{\mathrm{in}}})
-
(\ket{\psi_\rightarrow}^{\otimes n}
\otimes\ket{\Phi}_{F_{\mathrm{A}_n}^{\mathrm{out}}F_{\mathrm{B}_n}^{\mathrm{out}}})
\right\|_{1}  \\
& &~+\left\|
(\mathcal{S}^{\leftarrow}_{n}\otimes\mathds{1}_{R}^{\otimes n})
(\ket{\psi_\rightarrow}^{\otimes n}
\otimes\ket{\Phi}_{S_{\mathrm{A}_n}^{\mathrm{in}}S_{\mathrm{B}_n}^{\mathrm{in}}})
-
(
\ket{\psi_\rightleftarrows}^{\otimes n}
\otimes\ket{\Phi}_{S_{\mathrm{A}_n}^{\mathrm{out}}S_{\mathrm{B}_n}^{\mathrm{out}}}
)
\right\|_{1}\le\varepsilon_n,
\end{eqnarray*}
where $\ket{\Phi}_{\hat{E}_{\mathrm{A}_n}\hat{E}_{\mathrm{B}_n}}$
is an maximally entangled states
with Schmidt rank
$e^{\mathrm{out}}(\mathcal{F}^{\rightarrow}_{n})/e^{\mathrm{in}}(\mathcal{S}^{\leftarrow}_{n})$
and $\varepsilon_n=\varepsilon_n^{\rightarrow}+\varepsilon_n^{\leftarrow}$.
The first and second inequalities come
from the triangle property and the monotonicity of the trace distance~\cite{W13}.
Similarly,
if
$e^{\mathrm{out}}(\mathcal{F}^{\rightarrow}_{n})<e^{\mathrm{in}}(\mathcal{S}^{\leftarrow}_{n})$
then
\begin{eqnarray*}
& &\left\|
(\mathcal{E}^{\rightleftarrows}_n\otimes\mathds{1}_{R}^{\otimes n})
(\ket{\psi}^{\otimes n}
\otimes\ket{\Phi}_{F_{\mathrm{A}_n}^{\mathrm{in}}F_{\mathrm{B}_n}^{\mathrm{in}}}
\otimes\ket{\Phi}_{\tilde{E}_{\mathrm{A}_n}\tilde{E}_{\mathrm{B}_n}})
-
(
\ket{\psi_\rightleftarrows}^{\otimes n}
\otimes\ket{\Phi}_{S_{\mathrm{A}_n}^{\mathrm{out}}S_{\mathrm{B}_n}^{\mathrm{out}}}
)
\right\|_{1}  \\
&&\le
\left\|
((\mathcal{S}^{\leftarrow}_{n}\circ\tilde{\mathcal{F}}^{\rightarrow}_{n})
\otimes\mathds{1}_{R}^{\otimes n})
(\ket{\psi}^{\otimes n}
\otimes\ket{\Phi}_{F_{\mathrm{A}_n}^{\mathrm{in}}F_{\mathrm{B}_n}^{\mathrm{in}}}
\otimes\ket{\Phi}_{\tilde{E}_{\mathrm{A}_n}\tilde{E}_{\mathrm{B}_n}})
-(\mathcal{S}^{\leftarrow}_{n}\otimes\mathds{1}_{R}^{\otimes n})
(\ket{\psi_\rightarrow}^{\otimes n}
\otimes\ket{\Phi}_{F_{\mathrm{A}_n}^{\mathrm{out}}F_{\mathrm{B}_n}^{\mathrm{out}}}
\otimes\ket{\Phi}_{\tilde{E}_{\mathrm{A}_n}\tilde{E}_{\mathrm{B}_n}})
\right\|_{1}  \\
& &~+\left\|
(\mathcal{S}^{\leftarrow}_{n}\otimes\mathds{1}_{R}^{\otimes n})
(\ket{\psi_\rightarrow}^{\otimes n}
\otimes\ket{\Phi}_{F_{\mathrm{A}_n}^{\mathrm{out}}F_{\mathrm{B}_n}^{\mathrm{out}}}
\otimes\ket{\Phi}_{\tilde{E}_{\mathrm{A}_n}\tilde{E}_{\mathrm{B}_n}})
-
(
\ket{\psi_\rightleftarrows}^{\otimes n}
\otimes\ket{\Phi}_{S_{\mathrm{A}_n}^{\mathrm{out}}S_{\mathrm{B}_n}^{\mathrm{out}}}
)
\right\|_{1}  \\
&&\le
\left\|
(\mathcal{F}^{\rightarrow}_{n}\otimes\mathds{1}_{R}^{\otimes n})
(\ket{\psi}^{\otimes n}
\otimes\ket{\Phi}_{F_{\mathrm{A}_n}^{\mathrm{in}}F_{\mathrm{B}_n}^{\mathrm{in}}})
-
(\ket{\psi_\rightarrow}^{\otimes n}
\otimes\ket{\Phi}_{F_{\mathrm{A}_n}^{\mathrm{out}}F_{\mathrm{B}_n}^{\mathrm{out}}})
\right\|_{1}  \\
& &~+\left\|
(\mathcal{S}^{\leftarrow}_{n}\otimes\mathds{1}_{R}^{\otimes n})
(\ket{\psi_\rightarrow}^{\otimes n}
\otimes\ket{\Phi}_{S_{\mathrm{A}_n}^{\mathrm{in}}S_{\mathrm{B}_n}^{\mathrm{in}}})
-
(
\ket{\psi_\rightleftarrows}^{\otimes n}
\otimes\ket{\Phi}_{S_{\mathrm{A}_n}^{\mathrm{out}}S_{\mathrm{B}_n}^{\mathrm{out}}}
)
\right\|_{1}\le\varepsilon_n,
\end{eqnarray*}
where $\ket{\Phi}_{\tilde{E}_{\mathrm{A}_n}\tilde{E}_{\mathrm{B}_n}}$
is an maximally entangled states
with Schmidt rank
$e^{\mathrm{in}}(\mathcal{S}^{\leftarrow}_{n})/e^{\mathrm{out}}(\mathcal{F}^{\rightarrow}_{n})$.
It follows that a LOCC protocol $\mathcal{E}^{\rightleftarrows}_n$
is a state exchange with QSI of of $\ket{\psi}^{\otimes n}$ with error $\varepsilon_n$
together with
\begin{eqnarray*}
\lim_{n\to\infty}
\frac{\log e^{\mathrm{in}}(\mathcal{E}^{\rightleftarrows}_n)
-\log e^{\mathrm{out}}(\mathcal{E}^{\rightleftarrows}_n)}{n}
&=&\lim_{n\to\infty}
\frac{\log e^{\mathrm{in}}(\mathcal{F}^{\rightarrow}_{n})
      +\log e^{\mathrm{in}}(\mathcal{S}^{\leftarrow}_{n})
-\log e^{\mathrm{out}}(\mathcal{F}^{\rightarrow}_{n})
 -\log e^{\mathrm{out}}(\mathcal{S}^{\leftarrow}_{n})}{n} \\
&=&\lim_{n\to\infty}
\frac{\log e^{\mathrm{in}}(\mathcal{F}^{\rightarrow}_{n})
      -\log e^{\mathrm{out}}(\mathcal{F}^{\rightarrow}_{n})}{n}
      +
\lim_{n\to\infty}
\frac{\log e^{\mathrm{in}}(\mathcal{S}^{\leftarrow}_{n})
 -\log e^{\mathrm{out}}(\mathcal{S}^{\leftarrow}_{n})}{n} \\
&=&H(C_{\mathrm{B}}|A)+H(C_{\mathrm{A}}|BC_{\mathrm{B}}),
\end{eqnarray*}
where the first equality comes from the fact that
if $e^{\mathrm{out}}(\mathcal{F}^{\rightarrow}_{n})\ge e^{\mathrm{in}}(\mathcal{S}^{\leftarrow}_{n})$
then
\begin{equation*}
\log e^{\mathrm{in}}(\mathcal{E}^{\rightleftarrows}_n)
-\log e^{\mathrm{out}}(\mathcal{E}^{\rightleftarrows}_n)
=\log e^{\mathrm{in}}(\mathcal{F}^{\rightarrow}_{n})
  -\log e^{\mathrm{out}}(\mathcal{S}^{\leftarrow}_{n})
  - \log \dim \hat{E}_{A_n},
\end{equation*}
and if
$e^{\mathrm{out}}(\mathcal{F}^{\rightarrow}_{n})< e^{\mathrm{in}}(\mathcal{S}^{\leftarrow}_{n})$
then
\begin{equation*}
\log e^{\mathrm{in}}(\mathcal{E}^{\rightleftarrows}_n)
-\log e^{\mathrm{out}}(\mathcal{E}^{\rightleftarrows}_n)
= \log e^{\mathrm{in}}(\mathcal{F}^{\rightarrow}_{n})
  +\log \dim \tilde{E}_{A_n}
  -\log e^{\mathrm{out}}(\mathcal{S}^{\leftarrow}_{n}).
\end{equation*}
Since
\begin{equation*}
\lim_{n\to\infty} \varepsilon_n
=\lim_{n\to\infty}\varepsilon_n^{\rightarrow}
+\lim_{n\to\infty}\varepsilon_n^{\leftarrow}=0,
\end{equation*}
$u_{1}$ is an achievable rate of the state exchange with QSI of $\ket{\psi}$.

Moveover,
by relabeling Alice (Bob) with $\mathrm{Bob}'$ ($\mathrm{Alice}'$),
we obtain that $H(C_{\mathrm{A}}|B)+H(C_{\mathrm{B}}|AC_{\mathrm{A}})$ is also achievable.
Therefore, $e_{\mathrm{opt}} \le \min \{ u_{1}, u_{2} \}$.

\section{Proof of Corollary~\ref{cor:lb}} \label{app:proof_lb}

We note that
if $\log e^{\mathrm{in}}(\mathcal{E}_n)= n(H(C_\mathrm{A})+H(C_\mathrm{B}))$
then Alice and Bob can clearly perform the state exchange with QSI
by using the Schumacher compression~\cite{S95}
and the standard teleportation~\cite{BBCJPW93} on
the maximally entangled states with Schmidt rank $e^{\mathrm{in}}(\mathcal{E}_n)$.
Hence we may assume that $\log e^{\mathrm{in}}(\mathcal{E}_n)$
is not more than $n(H(C_\mathrm{A})+H(C_\mathrm{B}))$.

%

We now give a proof of Corollary~\ref{cor:lb}
which employs the continuity of the von Neumann entropy.

\begin{proof}
Let $e_\mathrm{r}$ be any achievable rate of the state exchange with QSI
of $\ket{\psi}\equiv\ket{\psi}_{AC_{\mathrm{A}}BC_{\mathrm{B}}R}$.
Then from the definition of the achievable entanglement rate,
there is a sequence $\{\mathcal{E}_n\}_{n\in\mathbb{N}}$ of state exchanges $\mathcal{E}_n$
with QSI of $\ket{\psi}^{\otimes n}$ with error $2\varepsilon_n$
such that
\begin{equation}\label{eq:norm_vNE}
\left\|
\rho_{X'Y'R^{\otimes n}}-
{\phi}_{X'Y'R^{\otimes n}}
\right\|_{1}
\le2\varepsilon_n,
\end{equation}
where $X'=A^{\otimes n}C_{\mathrm{B}}'^{\otimes n}E_{\mathrm{A}_n}^{\mathrm{out}}$ and
$Y'=B^{\otimes n}C_{\mathrm{A}}'^{\otimes n}E_{\mathrm{B}_n}^{\mathrm{out}}$,
\begin{eqnarray*}
\rho_{X'Y'R^{\otimes n}}&=&(\mathcal{E}_n\otimes\mathds{1}_{R^{\otimes n}})
\left({\psi}^{\otimes n}
\otimes{\Phi}_{E_{\mathrm{A}_n}^{\mathrm{in}}E_{\mathrm{B}_n}^{\mathrm{in}}}\right),\\
{\phi}_{X'Y'R^{\otimes n}}&=&{\psi'}^{\otimes n}
\otimes{\Phi}_{E_{\mathrm{A}_n}^{\mathrm{out}}E_{\mathrm{B}_n}^{\mathrm{out}}},
\end{eqnarray*}
${\Phi}_{E_{\mathrm{A}_n}^{\mathrm{in}}E_{\mathrm{B}_n}^{\mathrm{in}}}$
and ${\Phi}_{E_{\mathrm{A}_n}^{\mathrm{out}}E_{\mathrm{B}_n}^{\mathrm{out}}}$
are pure maximally entangled states
with Schmidt rank $e^{\mathrm{in}}(\mathcal{E}_n)$ and $e^{\mathrm{out}}(\mathcal{E}_n)$,
respectively,
$0\le \log e^{\mathrm{in}}(\mathcal{E}_n)\le n(H(C_\mathrm{A})+H(C_\mathrm{B}))$,
$\lim_{n\to\infty}
\frac{\log e^{\mathrm{in}}(\mathcal{E}_n)-\log e^{\mathrm{out}}(\mathcal{E}_n)}{n}=e_\mathrm{r}$,
and $\lim_{n\to\infty} \varepsilon_n=0$.
Then the monotonicity of the trace distance~\cite{W13} implies
\begin{eqnarray}
\left\| \rho_{X'}- \phi_{X'} \right\|_{1} &\le&2\varepsilon_n, \label{eq:Xprime}
\end{eqnarray}
where $\rho_{X'}=\Tr_{Y'R^{\otimes n}}[\rho_{X'Y'R^{\otimes n}}]$
and $\phi_{X'}=\Tr_{Y'R^{\otimes n}}[{\phi}_{X'Y'R^{\otimes n}}]$.

From the continuity of the von Neumann entropy~\cite{F73,A07} 
together with Eq.~(\ref{eq:Xprime}),
we obtain the following inequality:
\begin{equation} \label{eq:vNE_Cont}
\left| H(X')_{\rho_{X'}} - H(X')_{\phi_{X'}} \right|
\le
\varepsilon_n\log\dim(X')+h(\varepsilon_n),
\end{equation}
where $h(\cdot)$ is the binary entropy.
It follows that
\begin{eqnarray} \label{eq:main_vNE}
H(X')_{\rho_{X'}}
&\ge& H(X')_{\phi_{X'}}
-\varepsilon_n\log\dim(X')-h(\varepsilon_n) \nonumber \\
&=&nH(AC_{\mathrm{B}})+\log e^{\mathrm{out}}(\mathcal{E}_n)
-\varepsilon_n\left(n\log\dim(AC_{\mathrm{B}})
+\log e^{\mathrm{out}}(\mathcal{E}_n)\right) -h(\varepsilon_n),
\end{eqnarray}
and the von Neumann entropy $H(X')_{\rho_{X'}}$ is upper bounded as follows:
\begin{eqnarray} \label{eq:vNE_LB}
H(X')_{\rho_{X'}}
&\le&
E_d(X';Y'R^{\otimes n})_{\rho_{X'Y'R^{\otimes n}}}
+H(X'Y'R^{\otimes n})_{\rho_{X'Y'R^{\otimes n}}} \nonumber \\
&&\le
E_d(X;YR^{\otimes n})_{\ket{\psi}^{\otimes n}
\otimes\ket{\Phi}_{E_{\mathrm{A}_n}^{\mathrm{in}}E_{\mathrm{B}_n}^{\mathrm{in}}}}
+H(X'Y'R^{\otimes n})_{\rho_{X'Y'R^{\otimes n}}} \nonumber \\
&&\le
E_d(X;YR^{\otimes n})_{\ket{\psi}^{\otimes n}
\otimes\ket{\Phi}_{E_{\mathrm{A}_n}^{\mathrm{in}}E_{\mathrm{B}_n}^{\mathrm{in}}}}
+H(X'Y'R^{\otimes n})_{\ket{\phi}_{X'Y'R^{\otimes n}}}
+\varepsilon_n\log\dim(X'Y'R^{\otimes n})+h(\varepsilon_n),
\end{eqnarray}
where $E_d(X_1;X_2)$ is the distillable entanglement between $X_1$ and $X_2$ of a given state,
$X=A^{\otimes n}C_{\mathrm{A}}^{\otimes n}E_{\mathrm{A}_n}^{\mathrm{in}}$,
and $Y=B^{\otimes n}C_{\mathrm{B}}^{\otimes n}E_{\mathrm{B}_n}^{\mathrm{in}}$.
The first inequality comes from the hashing inequality~\cite{DW05}. 
Since the distillable entanglement is non-increasing under LOCC,
the second inequality holds.
The last inequality is obtained
from the continuity of the von Neumann entropy~\cite{F73,A07} 
together with Eq.~(\ref{eq:norm_vNE}).
Here,
$H(X'Y'R^{\otimes n})_{\ket{\phi}_{X'Y'R^{\otimes n}}}=0$,
since $\ket{\phi}_{X'Y'R^{\otimes n}}$ is pure.
Then the inequality in Eq.~(\ref{eq:main_vNE}) becomes
\begin{eqnarray*} \label{eq:main2_vNE}
E_d(X;YR^{\otimes n})_{\ket{\psi}^{\otimes n}
\otimes\ket{\Phi}_{E_{\mathrm{A}_n}^{\mathrm{in}}E_{\mathrm{B}_n}^{\mathrm{in}}}}
&\ge& nH(AC_{\mathrm{B}})+\log e^{\mathrm{out}}(\mathcal{E}_n) \\
&&-n\varepsilon_n\left(2n\log\dim(AC_{\mathrm{B}})
+n\log\dim(BC_{\mathrm{A}}R)
+\frac{3\log e^{\mathrm{out}}(\mathcal{E}_n)}{n}\right) -2h(\varepsilon_n).
\end{eqnarray*}
Thus
it follows that
\begin{equation*}
\frac{\log e^{\mathrm{in}}(\mathcal{E}_n)-\log e^{\mathrm{out}}(\mathcal{E}_n)}{n}
\ge
l_1
-\varepsilon_n\left(2\log\dim(AC_{\mathrm{B}})
+\log\dim(BC_{\mathrm{A}}R)
+\frac{3\log e^{\mathrm{out}}(\mathcal{E}_n)}{n}\right)
-\frac{2}{n}h(\varepsilon_n),
\end{equation*}
which implies that $e_\mathrm{r}\ge l_3$ as $n\rightarrow\infty$,
since
\begin{equation*}
0 \le\lim_{n\rightarrow\infty}\frac{\log e^{\mathrm{out}}(\mathcal{E}_n)}{n}
\le H(C_\mathrm{A}) + H(C_\mathrm{B}) + e_\mathrm{r}.
\label{eq:e_out_bounded}
\end{equation*}
Moreover,
the second lower bound $l_2$ can also be obtained in the same way
by replacing $H(X')$ in Eq.~(\ref{eq:vNE_Cont})
and $E_d(X';Y'R^{\otimes n})$ in Eq.~(\ref{eq:vNE_LB})
with
$H(Y')$ and $E_d(Y';X'R^{\otimes n})$,
respectively.
\end{proof}

%

\begin{Rem}
By employing the continuities of the quantum conditional entropy~\cite{W16}
and the quantum mutual information~\cite{W13}
instead of the von Neumann entropy
then we can get another two lower bounds,
$l_3=-H(AC_{\mathrm{B}}|BC_{\mathrm{A}})-H(AC_{\mathrm{A}})$
and $l_4=-H(BC_{\mathrm{A}}|AC_{\mathrm{B}})-H(BC_{\mathrm{B}})$,
on the optimal entanglement cost for the state exchange with QSI.
The lower bounds $l_3$ and $l_4$ are not tighter than $l_1$ and $l_2$, respectively.
\end{Rem}

\section{Examples} \label{app:example}
As mentioned earlier,
there are four QCMI conditions on the initial state of the state exchange with QSI
which give the exact optimal entanglement cost:
\begin{equation*}
I(R;C_\mathrm{A}|A)=0,
\quad I(R;C_\mathrm{A}|B)=0,
\quad I(R;C_\mathrm{B}|A)=0,
\quad I(R;C_\mathrm{B}|B)=0.
\end{equation*}
Let $S$ be the set of all pure states on the multipartite system $AC_{\mathrm{A}}BC_{\mathrm{B}}R$,
and define $S(X;Y|Z)$ as the intersection of $S$
and the set of all pure states which satisfy a condition $I(X;Y|Z)=0$.

We show that there are no inclusion relations among four sets
$S(R;C_\mathrm{A}|A)$, $S(R;C_\mathrm{A}|B)$, $S(R;C_\mathrm{B}|A)$, and $S(R;C_\mathrm{B}|B)$.
Consider the following state
\begin{equation*} 
\ket{\psi}_{AC_{\mathrm{A}}BC_{\mathrm{B}}R}
=\frac{1}{\sqrt{3}}(\ket{00000}+\ket{01100}+\ket{10011})_{AC_{\mathrm{A}}BC_{\mathrm{B}}R},
\end{equation*}
then we obtain
\begin{eqnarray*}
I(R;C_\mathrm{A}|A)
&=&(H(RA)-H(A))+(H(C_\mathrm{A}A)-H(RC_\mathrm{A}A))=0+0=0, \\
I(R;C_\mathrm{A}|B)
&=&(H(RB)-H(B))+H(C_\mathrm{A}B)-H(RC_\mathrm{A}B)
=\frac{2}{3}+H(C_\mathrm{A}B)-H(RC_\mathrm{A}B) \\
&\approx&0.66666+0.550048-0.918296>0, \\
I(R;C_\mathrm{B}|A)
&=&(H(RA)-H(A))+H(C_\mathrm{B}A)-H(RC_\mathrm{B}A)
=0+H(C_\mathrm{B}A)-H(RC_\mathrm{B}A) \\
&\approx&0.918296-0.550048>0, \\
I(R;C_\mathrm{B}|B)
&=&(H(RB)-H(B))+(H(C_\mathrm{B}B)-H(RC_\mathrm{B}B))
=\frac{2}{3}+0>0,
\end{eqnarray*}
since
\begin{eqnarray*}
&&H(RA)=H(A)=H(B)=H(RC_\mathrm{A}B)
=\frac{2}{3}\log_2\frac{3}{2}+\frac{1}{3}\log_2 3
\approx 0.918296, \\
&&H(C_\mathrm{B}B)=H(RC_\mathrm{B}B)=H(RB)
=\log_2 3
\approx 1.58496, \\
&&H(C_\mathrm{A}B)
=h\left(\frac{3-\sqrt{5}}{6}\right)
\approx 0.550048,
\end{eqnarray*}
where $h(\cdot)$ is the binary entropy.
Thus,
$\ket{\psi}\in S(R;C_\mathrm{A}|A)$,
$\ket{\psi}\notin S(R;C_\mathrm{A}|B)$,
$\ket{\psi}\notin S(R;C_\mathrm{B}|A)$,
and $\ket{\psi}\notin S(R;C_\mathrm{B}|B)$.
Moreover,
the other relations are easily shown by relabeling the subsystems of $\ket{\psi}$.
\vfill \clearpage
\end{widetext}

\end{document}